\documentclass[twocolumn]{jpsj3}
\usepackage{bm}

\title{Scalable Neutral Atom Quantum Computer with Interaction on Demand: 
Proposal for Selective Application of Two-Qubit Gate}

\author{
\name{Elham Hosseini \surname{Lapasar}}$^{1}$, 
\name{Kenichi \surname{Kasamatsu}}$^{1,2}$, 
\name{Yasushi \surname{Kondo}}$^{1,2}$, 
\name{Mikio \surname{Nakahara}}$^{1,2}$\thanks{E-mail address: nakahara@math.kindai.ac.jp}, 
\name{Tetsuo \surname{Ohmi}}$^{1}$, 
}
\inst{\address{$^{1}$Research Center for Quantum Computing, 
Interdisciplinary Graduate School of Science and Engineering, 
Kinki University, 3-4-1 Kowakae, Higashi-Osaka, 577-8502, Japan} \\
$^{2}$Department of Physics, Kinki University, 
3-4-1 Kowakae, Higashi-Osaka, 577-8502, Japan
}
\abst{
We propose a scalable neutral atom quantum computer with an on-demand
interaction through a {\it selective} two-qubit gate operation. 
Atoms are trapped by a lattice of near field Fresnel diffraction lights 
so that each trap captures a single atom.
One-qubit gate operation is implemented by a gate control laser beam 
which is applied to an individual atom. 
Two-qubit gate operation between an arbitrary pair of atoms
is implemented by sending these atoms
to a state-dependent optical lattice and making them collide 
so that a particular two-qubit state acquires a dynamical phase. 
We give numerical evaluations corresponding to these processes, 
from which we estimate the upper bound of a two-qubit gate operation 
time and corresponding gate fidelity. 
Our proposal is feasible within currently available 
technology developed in cold atom gas, MEMS, nanolithography, and various 
areas in optics. 
}

\kword{quantum computer, neutral atom, two-qubit gate}

\begin{document}
\maketitle

\section{Introduction}

Quantum computation employs a quantum mechanical system
as a computational resource \cite{roadmap,
NielsenChuang,nakaharaohmi}. It is expected that
a large scale quantum computer outperforms 
its classical counterpart exponentially
by making use of superposition states and entangled states. 
In spite of many proposals for potentially scalable quantum computers, 
most physical realizations so far accommodate qubits on the order of ten, at most.
One of the obstacles against scalability in the previous
proposals is the absence of controllable interaction between
an arbitrary pair of qubits. 
Small scale quantum computers are already demonstrated experimentally
in several physical systems, such as NMR, trapped ions, photons,
and neutral atoms. These physical systems have merits and demerits
and no single physical system to date satisfies all the DiVincenzo criteria,
the necessary conditions for a physical system to be a candidate
of a working quantum computer.\cite{ref:div}

A neutral atom quantum computer is one of the most promising 
candidates of a scalable quantum computer. 
Neutral atoms are believed to be robust against decoherence, another
obstacle against a physical realization of a working quantum computer, since
they are coupled very weakly to the environment. 
A one-qubit gate operation has been already demonstrated with a two-photon Raman 
transition,\cite{ref:1qgate,Jones} which is a well-established technique 
today. A two-qubit gate operation makes use of a cutting edge technique involving
ingenious manipulation of hyperfine-state-dependent optical
lattice potentials.\cite{Jaksch,ref:mandel1,ref:mandel2,Treutlein} 
A drawback of this two-qubit gate implementation
is that the gate acts on all the nearest neighbor pairs in the
optical lattice simultaneously. 
Hence a selective gate operation is yet to be realized, although some 
ideas utilizing optical tweezers has been proposed \cite{tweezer}. 

\begin{figure}
\begin{center}
\includegraphics[width=8.5cm]{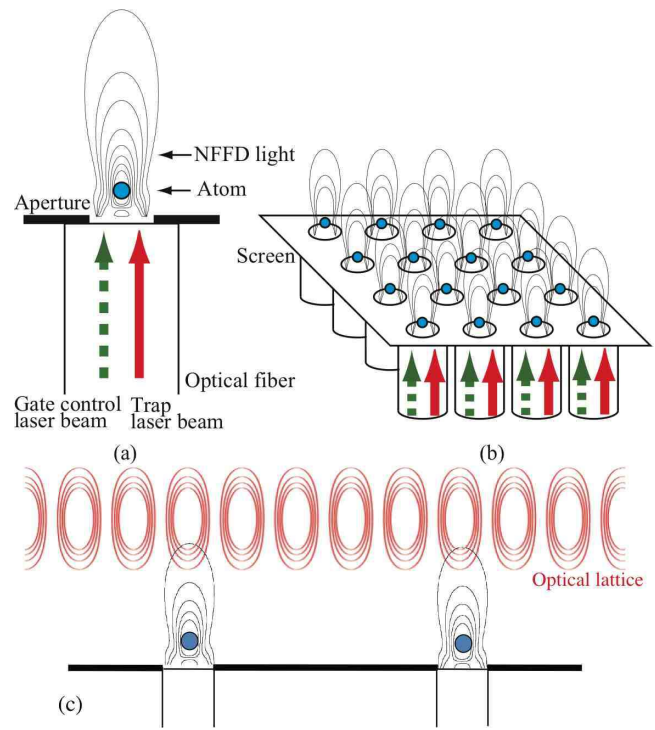}
\end{center}
\label{fig:concept}
\caption{(Color online)
(a) An atom trapped in an NFFD light. Red arrow (solid line) shows
the trap light while the green arrow (broken line) 
shows the laser light required
for one-qubit gate operations. (b) Example of an array of such traps. 
The trap laser and the
gate control laser are explicitly shown only for the first row. 
(c) A one-dimensional optical lattice is superimposed near the array 
of the NFFD traps. The minima of the optical lattice are situated so 
as to adjust the period of the array of the NFFD traps. }
\end{figure}
It is the purpose of the present paper to propose a new design of a
neutral atom quantum computer with an on-demand interaction,
which is potentially scalable up to a large number of qubits
within currently available technology \cite{Nakahara,Banchi}. 
The schematic illustration of our proposal is shown in Fig. 1. 
We punch an array of apertures in a thin substrate made of, e.g., silicon. 
An optical fiber is attached to each aperture, which lets in two laser
beams, one for trapping and the other for one-qubit gate control. 
Atoms are trapped by a near field Fresnel diffraction light (NFFD) \cite{Bandi}, 
which is produced by letting the trap laser light pass through 
an aperture with a diameter comparable to the wavelength of the laser, 
so that each trap holds a single atom. 
Turning on and off the trap laser
individually is required to implement a selective two-qubit gate 
as will be shown later.
Each lattice site is also equipped with its 
own gate control laser beam so that individual single-qubit gate operations 
can be applied on many qubits individually and simultaneously. 
A controlled two-qubit gate operation between an arbitrary pair of qubits 
is made possible by selectively transferring two atoms in a superimposed one-dimensional 
optical lattice and by employing the method demonstrated by Mandel {\it et al} \cite{ref:mandel2}. 
Here, we give detailed analysis of the selective two-qubit gate implementation 
according to our proposal. 

Technology required for physical implementation of the proposed 
quantum computer has already been developed in the areas of cold atom trap, 
scanning near field optical microscope (SNOM), nanolithography,
and quantum optics, among others, 
and we believe there should be no obstacles against its physical realization.

We first briefly review the design of a neutral atom quantum computer 
demonstrated by Mandel {\it et al.} \cite{ref:mandel2} in Sec. \ref{Mandelexp}. 
We point out problems inherent in their design. Individual atom trap by
making use of NFFD light is introduced in Sec. \ref{Banditrap}. 
Section \ref{imple} is devoted to one- and two-qubit gate implementations.
We separate the two-qubit gate operation into several steps and evaluate the execution time
and the corresponding fidelity of each step in Sec. \ref{executiontime} and Appendix\ref{appsimulation}. 
Conclusion is given in Sec. \ref{summary}.

\section{Brief Review of Neutral Atom Quantum Computer in Optical Lattice}\label{Mandelexp}
In this section, we look back the experiments by Mandel {\it et al.} 
\cite{Jaksch,ref:mandel1,ref:mandel2} to give 
a basic background on how to trap and manipulate 
atoms using laser beams and how to implement gate operation. 
We subsequently point out problems inherent in their design.

Suppose an atom is put in a laser beam with an oscillating electric field
$\bm{E}(\bm{x}, t) = {\rm Re} (\bm{E}_0(\bm{x}) e^{-i \omega_L t})$,
where $\omega_L$ is the laser frequency. It is assumed that $\omega_L$ is
close to some transition frequency $\omega_0 =E_e-E_g$ 
between two states $|g \rangle$
and $|e \rangle$ of the atom. The interaction between the electric field
and the dipole moment of the atom introduces an interaction Hamiltonian
\begin{equation}
H_i = -\frac{1}{2} (\bm{E}_0 \cdot \bm{d}) (e^{-i \omega_L t}
+ e^{i \omega_L t}),
\end{equation}
where $\bm{d}$ is the dipole moment operator of the atom.
This interaction introduces an effective potential of the form
\begin{equation}
V(\bm{x}) = \frac{\hbar |\Omega_{eg}(\bm{x})|^2}
{4\Delta_{eg}}
\end{equation}
for an atom in the ground state, which is called the AC Stark shift.
Here $\Omega_{eg}=\langle e|\bm{d}|g\rangle \cdot \bm{E}_0(\bm{x})/\hbar$, while
$\Delta_{eg} = \omega_L-\omega_0$ is the detuning. 
We have ignored the small natural line width of the excited state. 
In case $\Delta_{eg}>0$
(blue-detuned laser), $V(\bm{x})$ is positive and a region with large
$V(\bm{x})$ works as a repulsive potential. In contrast, if $\Delta_{eg}<0$
(red-detuned laser), $V(\bm{x})$ is negative and a region with large
$|V(\bm{x})|$ works as an attractive potential. We use mainly a red-detuned laser
in the following proposals.

It is possible to confine neutral atoms 
by introducing a pair of counterpropagating laser beams with the same
frequency, amplitude and polarization along the $x$-axis. 
These beams produce a standing wave potential, called an ``optical lattice" 
of the form
\begin{equation}
V_{\rm OL}({\bm x}) = - V_{0} \cos^2 (k_{\rm OL} x) e^{-2(y^2+z^2)/w^2}, 
\label{opticallatticeeq}
\end{equation}
where $k_{\rm OL}$ is the wave number of the laser and $w$ is the waist size,
which is on the order of the laser wavelength. 
It is possible to construct a three-dimensional lattice by 
adding two sets of counterpropagating
beams along the $y$- and the $z$-axes. 
Note that the lattice constant is always $\lambda_{\rm OL}/2$,
where $\lambda_{\rm OL}$ is the wavelength of the laser.

Mandel {\it et al.} trapped atoms in such an optical lattice \cite{ref:mandel2}.
They used the Rabi oscillation to implement one-qubit gates, in which
a microwave (MW) field was applied. 
They have chosen qubit basis vectors 
$|0 \rangle =|F=1, m_F=-1 \rangle$ and $|1 \rangle =|F=2, m_F=-2 \rangle$
to be in harmony with the one-qubit gate operation making use of the
Rabi oscillation. 

For two-qubit gate operations, they introduced time-dependent
polarization in the counterpropagating laser beams as
\begin{equation}
\begin{array}{c}
\bm{E}_+(\bm{x}) = e^{ik_{\rm OL}x}(\hat{\bm{z}} \cos \theta +
\hat{\bm{y}} \sin \theta),
\vspace{.2cm}\\
\bm{E}_-(\bm{x}) = e^{-ik_{\rm OL}x}(\hat{\bm{z}} \cos \theta -
\hat{\bm{y}} \sin \theta).
\end{array}
\end{equation}
These counterpropagating laser beams produce an optical potential
of the form
\begin{equation}
\bm{E}_+(\bm{x})+\bm{E}_-(\bm{x}) \propto \sigma^+ \cos (k_{\rm OL}x-\theta)
+ \sigma^- \cos (k_{\rm OL}x+\theta),
\end{equation}
where $\sigma^+$ ($\sigma^-$) denotes counterclockwise (clockwise)
circular polarization. Note that the first component ($\propto \sigma^+$)
moves along the $x$-axis as $\theta$ is increased, while the
second component ($\propto \sigma^-$) moves along the $-x$-axis under
this change \cite{ref:mandel1}.
The component $\sigma^+$ introduces transitions between
fine structures; $nS_{1/2}\ (m_J=-1/2) \to nP_{1/2}\ (m_J=1/2)$, 
$nS_{1/2}\ (m_J=-1/2) \to nP_{3/2}\ (m_J= 1/2)$, and $nS_{1/2}\ (m_J=
1/2) \to nP_{3/2}\ (m_J=3/2)$. The transitions from $nS_{1/2}$ to
$nP_{3/2}$ are red-detuned, while the transition from $nS_{1/2}$ to
$nP_{1/2}$ is blue-detuned if $\omega_L$ is chosen between the transition
frequencies of $nS_{1/2}\ (m_J=-1/2) \to nP_{3/2}\ (m_J=1/2)$ and
$nS_{1/2}\ (m_J=-1/2) \to nP_{1/2}\ (m_J=1/2)$. Then by adjusting
$\omega_L$ properly, it is possible to cancel the attractive potential
and the repulsive potential associated with these transitions.
The net contribution of the $\sigma^+$ laser beam in this case is an attractive
potential for an atom in the state $nS_{1/2}\ (m_J= 1/2)$. We denote this
potential as $V_+(\bm{x}) \propto \cos^2 (k_{\rm OL}x - \theta) $ in the following.
Similarly, the $\sigma^-$ component introduces a net attractive potential
$V_-(\bm{x})  \propto \cos^2 (k_{\rm OL}x + \theta)$, through the transition $nS_{1/2}\ (m_J=-1/2)
\to nP_{3/2}\ (mJ=-3/2)$ on an atom in the state $nS_{1/2}\ (m_J=-1/2)$. 

Mandel {\it et al.} \cite{ref:mandel2} took advantage of these state-dependent
potentials to implement a two-qubit gate. 
By applying the Walsh-Hadamard gate on $|0 \rangle_i |0 \rangle_{i+1}$,
one generates a tensor product state 
$(|0 \rangle +|1 \rangle)_i(|0 \rangle + |1 \rangle)_{i+1}/2$. 
The potentials acting on the states 
$|0 \rangle$ and $|1 \rangle$ are evaluated as
\begin{equation}
\begin{array}{c}
V_{|0 \rangle}(\bm{x}) = \dfrac{3}{4} V_+(\bm{x}) + \dfrac{1}{4} V_-(\bm{x})
\vspace{.2cm}\\
V_{|1 \rangle}(\bm{x}) = V_-(\bm{x}).
\end{array}
\label{statepotMan}
\end{equation}
By decreasing the phase $\theta$, the state $|0 \rangle$ 
moves with dominating $V_+$ toward $+x$-direction, while $|1 \rangle$ moves
with $V_-$ toward $-x$-direction. Thus it is possible to make $|0 
\rangle_i$ and $|1 \rangle_{i+1}$ collide between the two lattice
points. If they are kept in the common potential well during $t_{\rm hold}$,
the subspace $|0 \rangle_i|1 \rangle_{i+1}$ obtains a dynamical phase
$e^{-i U_{\rm int} t_{\rm hold}}$ \cite{Jaksch}, where $U_{\rm int}$ is the on-site repulsive potential 
energy. After these two components spend $t_{\rm hold}$
in the potential well, they are brought back to the initial lattice point
by reversing $\theta$, which results in the two-qubit gate operation
\begin{eqnarray}
|0 \rangle_i |1 \rangle_{i+1}
\to
\frac{1}{2} \biggl( |0 \rangle_i |0 \rangle_{i+1} +e^{-i U_{\rm int} t_{\rm hold}}
|0 \rangle_i |1 \rangle_{i+1} \nonumber\\
+
|1 \rangle_i |0 \rangle_{i+1} +|1 \rangle_i |1 \rangle_{i+1} \biggr).
\end{eqnarray}
It should be noted, however, that the state dependent potentials act on
all the pairs of the atoms and selective operation of the two-qubit
gate on a particular pair is impossible. Their proposal may be applicable to
generate a highly entangled state for a cluster state quantum computing, 
although it is not applicable for a circuit model quantum computation.

In the following sections, we propose implementations of a neutral
atom quantum computer, which overcome these difficulties.

\section{Near Field Fresnel Diffraction Trap}\label{Banditrap}
Suppose there is an array of apertures in a thin substrate. We consider
a square lattice of apertures for definiteness. The substrate is
made of a thin silicon, for example. Silicon has a very good
thermal conductivity, comparable to a metal, and local heating
due to a laser is expected to diffuse quickly. It is possible
to maintain the substrate in a low temperature if the perimeter
of the substrate is in contact with a refrigerator, which suppresses
thermal excitation of an atom. 

Bandi {\it et al.} \cite{Bandi} proposed to trap an atom with a microtrap 
employing NFFD light. 
Atoms are trapped in an array of NFFD traps, each of which traps a single atom.
An NFFD light is produced if a plane wave with the wave length $\lambda_{F}$ 
is incident to a screen with an aperture of the radius $a \geq \lambda_{F}$.
The ratio $N_F = a/\lambda_{F} \geq 1$ is called the Fresnel number. 
The trap potential is evaluated by applying the 
Rayleigh-Sommerfeld formula as
\begin{equation}
U_F(\bm{x}) = -U_0 \frac{|\mathcal{E}(\bm{x})|^2}{E_0^2}, \label{eq:rs}
\end{equation} 
where
\begin{equation}
\mathcal{E}(\bm{x}) = \frac{E_0}{2\pi} \iint \frac{e^{ik_{\rm F} r}}{r} \frac{z}{r}
\left(\frac{1}{r}-ik_{\rm F} \right) dx'dy'
\end{equation}
and 
\begin{equation}
U_0 = \frac{3}{8} \frac{\Gamma_e}{|\Delta_{eg}|} \frac{E_0^2}{k_{\rm F}^3}
\end{equation}
with the distance $r$ between ${\bm x}=(x,y,z)$
and ${\bm x}'=(x', y', 0)$ in the plane. 
Here, $\Gamma_e$ is half of the spontaneous decay rate, 
$E_0$ the amplitude of the incoming plane wave, $k_{\rm F}=2 \pi/\lambda_{F}$ 
its wave number, the detuning $\Delta_{eg}$ is negative (red-detuned),
and the plane wave is incident to a screen from the $-z$-axis.
The integral is over the aperture region $(x'^2+y'^2 \leq a^2)$.

\begin{figure}
\begin{center}
\includegraphics[width=8cm]{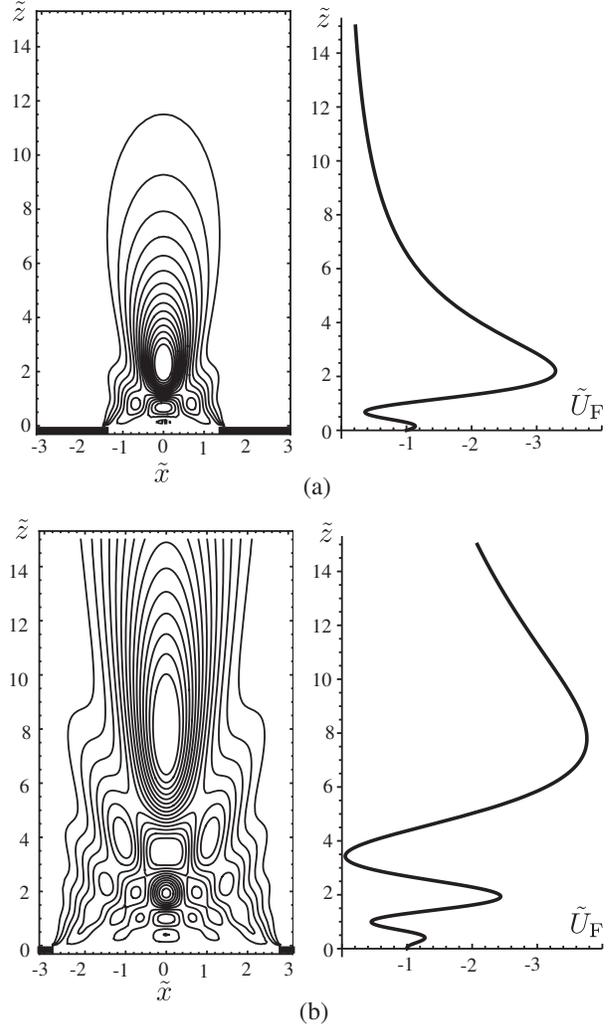}
\end{center}
\caption{Potential profiles of the NFFD trap for (a) $a/\lambda_{\rm F}= 1.5$,
and (b) $a/\lambda_{\rm F}=2.8$, where $a$ is the aperture
radius and $\lambda_{\rm F}$ the wavelength of the trap laser beam.
The left panel is the contour plot while the right panel shows the
potential profile along the $z$-axis with $x=y=0$. The length and the potential 
energy are normalized by $\lambda_{F}$ and $U_{0}$, respectively.}
\label{fig:nffd}
\end{figure}
\begin{figure}
\begin{center}
\includegraphics[width=6cm]{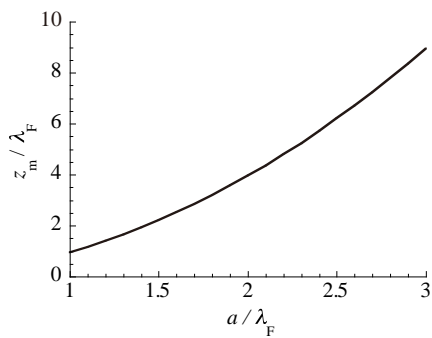}
\end{center}
\caption{$z$-coordinate of the potential minimum ($z_m$) as a function
of the aperture size $a/\lambda_{\rm F}$.}
\label{fig:zm}
\end{figure}

The aperture size is on the order of the wave length $\lambda_{F}$ so that 
the distance between the minimum of the potential energy 
and the substrate is on the order of the wavelength.
Figure \ref{fig:nffd} shows the trap potential profiles for two
aperture radii, which is obtained by evaluating Eq.~(\ref{eq:rs}) numerically.
It is clear from Fig. \ref{fig:nffd}. that the distance between the atom and the
substrate is controllable by manipulating the aperture size $a$. 
Figure \ref{fig:zm} shows the $z$-coordinate of the potential minimum ($z_{m}$) 
as a function of the aperture size $a$.
We take advantage of this fact in our implementation of a two-qubit
gate later. 

The aperture size may be controlled by employing 
a small scale liquid crystal technology called 
Spatial Light Modulator (SLM) with Liquid Crystal on Silicon (LCOS)
technology \cite{ref:hamamatsu,ref:lcos}. 
An array of shutters made of liquid crystal may be fabricated 
on a silicon wafer with LCOS-SLM. 
The current operation time
of the SLM is reported to be sub msec, which is comparable to the operation
times of the other steps in our gate implementation discussed below. 
Alternatively, well-established Micro-Electro-Mechanical Systems (MEMS) 
technology may be employed to control the aperture radius. 
An MEMS shutter for a display is already in a mass production stage
\cite{mems1} and it may be applied to demonstrate our proposal. 
The on (off) switching time of the shutter is 54 (36)~$\mu$s \cite{mems1}
and we expect 
even faster switching time by optimizing its structure for our 
purpose \cite{mems2}.


It is clear by construction that we have freedom in the
choice of the lattice constant and the lattice shape. We may even
arrange the traps in an irregular form if necessary.
This is a remarkable
new feature, which a conventional two-dimensional optical lattice 
cannot provide with; 
an optical lattice is always regular and its lattice constant
is on the order of the wavelength of the laser beams, where the 
lattice constant can be tuned to some extent by changing the 
angle of two lasers \cite{Peil}.

\section{Quantum Gate Implementations}\label{imple}
For a quantum system to be a candidate of a general purpose quantum computer,
it is necessary for the system to be able to implement a universal
set of quantum
gates, that is, the set of one-qubit gates and almost any two-qubit gate
\cite{Barenco}. Now we explain how these gates are realized in our proposal. 

In the following we take two hyperfine states to represent two-qubit basis 
vectors, which we denote $|0 \rangle$ and $|1 \rangle$. 
The hyperfine states must be consistent with two-qubit gate
operations and we take, for definiteness,
$|0 \rangle =|F=1, m_F=1 \rangle$ and $|1 \rangle =|F=2, m_F=1 \rangle$
throughout this paper.

\subsection{One-Qubit Gates}
Let us assume, for definiteness, the NFFD trap potential is strong enough 
so that there is only one atom in each trap. Along with the NFFD trap light,
there is a gate control laser beam propagating through
the fiber, which is depicted in a broken line in Fig. \ref{fig:concept}. 

One-qubit gate is implemented by making use of the two-photon Raman transition
\cite{nakaharaohmi}, which is already demonstrated \cite{ref:1qgate,Jones}. 
Let $E_0$ and $E_1$ be the energy eigenvalues of
the states $|0 \rangle$ and $|1\rangle$, respectively, and
let $E_e$ be the energy eigenvalue of an auxiliary excited state $|e\rangle$
necessary for the Raman transition. Suppose that a laser beam with the
frequency $\omega_L$ has been applied to the atom. Let 
$\Delta = \omega_L-(E_e-E_0)/\hbar$ be the detuning and
$\Omega_i$ be the Rabi oscillation frequency between the state $|i \rangle
\ (i=0, 1)$ and $|e \rangle$.
Then, under the assumptions $|\Delta| \gg (E_1-E_0)/\hbar, \Omega_i^2/|
\Delta|$, we obtain the effective Hamiltonian
\begin{equation}
H_1 = \frac{1}{2} \epsilon \sigma_z - \frac{\Omega_0 \Omega_1}{4 \Delta}
\sigma_x,
\end{equation}
where 
$$
\epsilon = E_1-E_0 + \frac{\Omega_1^2-\Omega_0^2}{4 \Delta}.
$$
Note that $H_1$ generates all the elements of SU(2) since there are
two $\mathfrak{su}(2)$ generators $\sigma_{x,z}$ in the Hamiltonian and 
their coefficients are controllable.

\subsection{Two-Qubit Gates}\label{twoqubitgate}
As illustrated in Fig. \ref{fig:twobit}, our proposal for selective two-qubit 
gate operation consists of 7 steps: 
\begin{figure}
\begin{center}
\includegraphics[width=8cm]{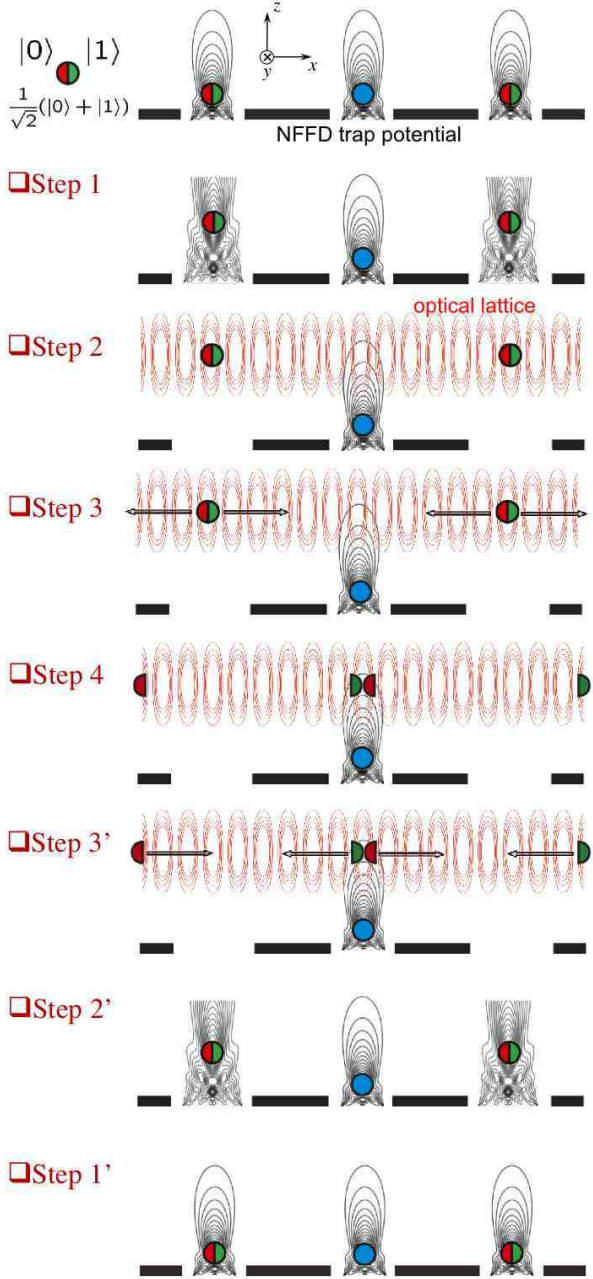}
\end{center}
\caption{(Color online) Schematic diagram of a two-qubit gate operation. 
In the top panel, the left and the right atoms are acted by the Walsh-Hadamard
gate in advance so that each of them is in the superposition of
$|0 \rangle$ (red semicircle) and $|1 \rangle$ (green semicircle). 
Here, the black contours show the individual NFFD potential and bold black
segments show the screen. 
Step 1: By enlarging the aperture size, the minima of the selected traps 
are moved away from the substrate. 
Step 2: A one-dimensional optical lattice along the $x$-axis, depicted by the thin red contours, 
is superimposed with turning off adiabatically the selected NFFD traps. 
Two atoms in the superposition of $|0 \rangle$ and $|1 \rangle$ 
are left in the state-dependent optical lattice. 
Step 3: The polarizations of the 
counterpropagating laser beams are tilted by $\pm \theta$ so that
the $|0 \rangle$ and $|1 \rangle$ components of two atoms 
move in the opposite directions. 
Step 4: The collision of $|0 \rangle$ of one
atom and $|1 \rangle$ of the other gives a dynamical phase. 
Steps 3', 2', and 1' are the reverse process of steps 3, 2, and 1, respectively.}
\label{fig:twobit}
\end{figure}

\begin{enumerate}
\item[1] 
We choose two trapped atoms to be operated by the gate. 
Each atom is made into a superposition
state $(|0 \rangle + |1 \rangle)/\sqrt{2}$ by a gate control
laser beam attached to each aperture. Then, 
these atoms are transferred from the vicinity
of the substrate to space coordinates, where a one-dimensional optical
lattice will be introduced later, by enlarging the aperture radius.
The minima of the optical lattice will be situated at these space coordinates. 

\item[2] While the atoms stay at the points, the NFFD lasers of the two atoms 
are gradually turned off while a pair of counterpropagating laser
beams to form an optical lattice is gradually turned on. 

\item[3] Now the two atoms are trapped in the optical lattice. 
As described in Sec. \ref{Mandelexp},
the polarizations of the pair of the counterpropagating laser beams
are rotated in opposite directions so that the qubit state $|0 \rangle$ 
is transferred toward positive $x$-direction, while $|1 \rangle$ is
transferred toward negative direction. Therefore, it is possible to 
collide $|0 \rangle$ of one atom and $|1 \rangle$ of the other atom. 

\item[4] Now, component $|0\rangle$ of one atom and $|1 \rangle$ of the
other are trapped in the same potential well in the optical lattice.
They interact with each other for a duration $t_{\rm hold}$ so that
this particular state $|0 \rangle |1 \rangle$ acquires an extra phase $U_{\rm int} t_{\rm hold}$
compared to other components $|0 \rangle|0 \rangle, |1 \rangle
|0 \rangle$, and $|1 \rangle|1 \rangle$. Here 
$U_{\rm int} = (4\pi \hbar^2 a_s/m) \int |\psi|^4 d\bm{x}$ 
is the on-site interaction energy with the atomic mass $m$ and 
the $s$-wave scattering length $a_s$.
When $t_{\rm hold}$ satisfies the condition 
$U_{\rm int} t_{\rm hold}= \pi$, we obtain a 
nonlocal two-qubit gate 
$|00 \rangle \langle 11|-|01 \rangle \langle 01|
+|10 \rangle \langle 10|+|11 \rangle \langle 11|$.
Application of further one-qubit gates implements a CNOT gate or a CZ gate.

\item[3'] After acquiring the phase, two atoms are separated
along the optical lattice by an inverse process of Step 3.

\item[2'] Atoms are transferred from the optical lattice to
the NFFD trap by an inverse process of Step 2.

\item[1'] Aperture radii of NFFD lasers are reduced so that
the atoms are transferred back to the vicinity of the substrate
by an inverse process of Step 1. Now the gate operation is
completed.

\end{enumerate}

\section{Execution Time of Two-Qubit Gate} \label{executiontime}
In this section, we analyze all steps of selective two-qubit gate operation 
in detail one by one to estimate the execution time. 
While shorter execution time is preferred, the fidelity should be kept close to 1
during the gate operation.  
The fidelity is mostly bounded by adiabaticity requirement. 
Here, we give an estimate of the execution time, 
which is limited by the adiabatic condition. 
Detailed numerical simulations, using the time-dependent 
Schr\"{o}dinger equation with experimentally realistic parameters, for each
step will be described in Appendix A. 
 
\subsection{Step 1}
\begin{figure}
\begin{center}
\includegraphics[width=8cm]{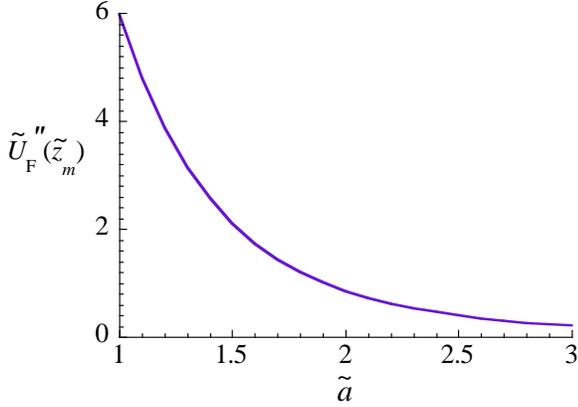}
\end{center}
\caption{(Color online) Curvature of the potential along the $z$-axis 
at the potential minimum as a function of the aperture radius. 
The length and the energy are normalized by $\lambda_{F}$ and $U_{0}$, respectively.}
\label{fig:curvature}
\end{figure}

Suppose that an atom is transferred away from the vicinity
of the substrate (from $z_m = z_{\rm ini}$ to $z_{\rm fin}$) by increasing the aperture 
radius $a$, e.g. from (a) to (b) of Fig. \ref{fig:nffd}. 
The distance $z_{\rm fin}-z_{\rm ini}$ may be a few times of $\lambda_{F}$.
As seen from Fig. \ref{fig:nffd}, the NFFD trap potential in this parameter range
is shallow along the $z$-axis and steep in the xy-plane, which means
that the curvature along the $z$-axis dominates the adiabaticity condition. 
The angular frequency along the $z$-axis around the minimum $z=z_m$ is
\begin{equation}
\omega_z = \sqrt{\left. \frac{1}{m} \frac{\partial^2 U_{\rm F}}{\partial z^2}\right|_{z=z_m}}
= \sqrt{\left. 2 \alpha \frac{\partial^2 \tilde{U}_{\rm F}}{\partial \tilde{z}^2} \right|_{\tilde{z}=\tilde{z}_m}} \frac{1}{\tau_{\rm F}},
\end{equation}
where we put tildes for dimensionless variables scaled by the wavelength $\lambda_{\rm F}$ 
and the potential amplitude $U_{0}$ of the NFFD trap Eq. (\ref{eq:rs}); 
$\tau_{\rm F} = \hbar/U_{0}$ is the characteristic time scale for NFFD trap and 
$\alpha = \hbar^2/2mU_{0}\lambda_{\rm F}^2$. The dimensionless curvature 
$\tilde{U}_{\rm F}'' (\tilde{z}_{m}) \equiv \partial^2 \tilde{U}_{\rm F}/{\partial \tilde{z}^2} 
|_{\tilde{z}=\tilde{z}_{m}}$ as a function of the aperture size $a$ 
is shown in Fig. \ref{fig:curvature}. 

Since $\omega_{z}$ is a monotonically decreasing function of $a$, 
the characteristic time $\tau_{\rm ad}^{(1)}$ for an adiabatic change 
can be estimated from the condition 
\begin{equation}
\frac{1}{2} m \left( \frac{z_{\rm fin} - z_{\rm ini}}{\tau_{\rm ad}^{(1)}} \right)^2 \sim \hbar \omega_{z}(z_{m} = z_{\rm fin} ),
\end{equation}
which gives
\begin{equation}\label{eq:tad1}
\tau_{\rm ad}^{(1)} \sim 
\frac{\tilde{z}_{\rm fin} - \tilde{z}_{\rm ini}}{\alpha^{\frac{3}{4}}  \tilde{U}_{\rm F}''(\tilde{z}_{\rm fin})^{\frac{1}{4}} } \tau_{\rm F}
\end{equation}
Here we have taken the ``worst'' curvature at the final value of $a$, i.e., $z_m=z_{\rm fin}$. 
For the typical parameter values for a realistic NFFD trap (see Appendix\ref{appstep1}) 
we can see $\alpha \ll 1$, because the depth of the NFFD trap is very deep. 
We expect, from this estimation, that the transfer of an atom from
the initial position to the optical lattice potential minimum
will be bounded in time by 
$\tau_{\rm ad}^{(1)} \sim 10^{3-4} \tau_{\rm F} \sim 10^{2-3} \mu{\rm s}$.

\subsection{Step 2}
Now that the atoms are transferred to the right space points, 
their wavefunctions must be transformed from the ground states of the NFFD traps 
to those of the optical lattice adiabatically. 
It is important to note that the centers 
of the NFFD traps and those of the optical lattice potential must be sitting
at the same space points for a smooth transfer of atoms. 
The adiabaticity in this case is limited by the
smallest potential curvature, that is, the curvature of
$V_{\rm OL}$ (Eq. (\ref{opticallatticeeq})) along the {\it radial} ($y$ or $z$) direction, 
because the typical laser waist satisfies $w \geq 4 \lambda_{\rm OL}$. 
The separation between 
the ground state and the first excited state energies $\hbar \omega_z$
gives a time scale $\tau_{\rm ad}^{(2)} \sim 2\pi/\omega_{z}$, where the angular frequency 
is defined as 
\begin{equation}
m \omega_z^2 = \frac{\partial^2 V_{\rm OL}}{\partial z^2} \biggr|_{z=0}=
\frac{4 V_0}{w^2} .
\end{equation}
Using the typical time scale $\tau_{\rm OL} = \hbar / E_{\rm r}$ with the recoil energy 
$E_{\rm r} = \hbar^{2} k_{\rm OL}^2 / 2 m$ of the optical lattice, 
the time scale is expressed as 
 \begin{equation}
\tau_{\rm ad}^{(2)} \sim w k_{\rm OL} \sqrt{\frac{E_{\rm r}}{V_{0}}} \tau_{\rm OL},
\end{equation}
which gives a time scale $\tau_{\rm ad}^{(2)} \sim 10^{1-2} \tau_{\rm OL} \sim 10^{2-3} \mu$s 
for typical parameter values (Appendix \ref{appstep2}). 

\subsection{Step 3}\label{step3estimate}
Now, the atoms are trapped in the state-dependent optical lattice. 
According to our construction, the length between two nearest-neighbor 
apertures in the substrate is larger than a few times of the wavelength 
of laser beams. 
Then it is difficult to load two atoms in the nearest neighbor minima of the 
optical lattice, because the lattice constant is $\lambda_{\rm OL}/2$. 
To collide the atoms, we must move the $| 0 \rangle$ ($| 1 \rangle$) atom 
right (left) by $n \times \lambda_{\rm OL}/2$, if the initial separation 
between the two atoms is $2n \times \lambda_{\rm OL}/2$; 
$n$ may be more than 5. 

The rough estimate of the execution time can be obtained by 
considering the adiabatic transport by one lattice constant 
$\lambda_{\rm OL}/2$ along the $x$-axis in the 
usual optical lattice Eq. (\ref{opticallatticeeq}). The adiabaticity condition is 
then given by $m(\lambda_{\rm OL}/2 T_{\rm OL})^2 / 2  \sim \hbar \omega_{x} = \hbar k_{\rm OL} \sqrt{2V_0/m}$. 
Thus, the adiabatic time scale for the motion 
with $n$ lattice constant is estimated as 
\begin{equation}
\tau_{\rm ad}^{(3)} \simeq n \times T_{\rm OL} 
\sim n \left( \frac{E_r}{V_0} \right)^{\frac{1}{4}}  \tau_{\rm OL}, 
\end{equation}
where $\tau_{\rm OL}$ has been defined in Step 2. 

In our choice of the hyperfine states for $|0 \rangle$ and $|1 \rangle$, 
the state-dependent lattice potentials are given by Eq. (\ref{statedependpot}) 
in Appendix\ref{appsimulation}. They are superpositions of two counterpropagating
plane waves $V_{\pm}$ and there is an interference between them. 
Then, the simple adiabatic argument cannot apply to this situation. 
Actually, as shown in Appendix\ref{appsimulation}, the fidelity is 
an oscillating function of the operation time for a given $n$; 
in other words, the adiabaticity time is not a monotonous function of $n$. 
We find through the numerical simulation that the execution time is 
roughly a few tens of the typical time scale $\tau_{\rm OL}$. 

\subsection{Step 4}
Let $\psi_0(\bm{x})$ be the ground state wave function of an atom
in one of the optical lattice potential wells . 
If two atoms in the ground state are put in the same potential
well, the interaction energy is given by 
\begin{equation}
U_{\rm int} 
 = \frac{2a_s E_{\rm r}}{\pi \lambda_{\rm OL}} \int \psi_0^4(\bm{x}) d\bm{x}. 
\end{equation}
To acquire a phase $\pi$ necessary for the controlled-$Z$ gate, for example,
these atoms must be kept in the same potential well for $t_{\rm hold} = \pi \hbar/U_{\rm int}$. 

For the $s$-wave scattering length $a_s=5.19$~nm between 
two $^{87}$Rb atoms in $|F=1, m_F=1
\rangle$ and $|F=2, m_F=1\rangle$ \cite{ref:sscat} 
and typical parameters of the optical lattice (see Appendix\ref{appsimulation}), 
we find $\tau_{\rm hold}$ of a few msec, which is the largest 
execution time among all the steps. 
This step does not involve any nonadiabatic process and we believe
the operation is executed with a high fidelity by fine tuning the
holding time $t_{\rm hold}$.




\subsection{Step 3', 2' and 1'}
It is clear that Step $n$' is an inverse process of Step $n$ and
time required for and the fidelity of the inverse process are the
same as those for a forward process. The mathematical proof 
is described in Appendix\ref{backward}.

\subsection{Total execution time and fidelity}
We have estimated the time required for each step, 
under the assumption that each step gives a fidelity close to 1. 
The overall execution time of a two-qubit gate is given by 
$T_{\rm overall} \simeq 2 \left( \tau_{\rm ad}^{(1)} 
+\tau_{\rm ad}^{(2)} + \tau_{\rm ad}^{(3)} \right) 
+ \tau_{\rm hold}$. 
The rough estimate under the adiabaticity condition gives $\sim$ 10 ms. 
Through the numerical simulation of the Schr\"{o}dinger equation with realistic parameters, 
as shown in Appendix\ref{appsimulation}, we obtain $7.87$ msec. 

The overall fidelity is estimated as follows. Each of Steps 1, 2, 3, 3', 2' 
and 1' involves two independent processes. For example, sending an atom from
the vicinity of the substrate to the space point in Step 1 is realized with
the fidelity 0.99 for each atom. Since this step involves two atoms, the
fidelity associated with this step must be $0.99^2$. Thus the overall fidelity
is $0.99^{12} \sim 0.886$, where we assumed that the interaction time 
is tunable so that Step 4 gives a fidelity arbitrarily close to 1.
The fidelity may be improved by spending more time for each step. 

Execution time may be shortened by applying stronger
laser fields, because the separation between neighboring energy levels
is enlarged by this, which makes fast adiabatic operation possible. Besides, stronger laser
fields suppress tunneling of an atom in the optical lattice
to the adjacent potential minima. The ground state wave function is 
squeezed in a stronger lattice potential well, leading to a
larger value of $\int \psi^4 d\bm{x}$. Then, Step 4 is also executed in a shorter
time. To make this statement more concrete, 
let us make a harmonic approximation to the
potential well of the optical lattice. One can see that $t_{\rm hold}$
is reduced by a factor $V_0^{-3/4}$ if $V_0$ is made larger; for example,
$t_{\rm hold}$ is multiplied by $0.178$ if $V_0$ becomes 10 times larger.
We expect that overall execution time can be 
reduced by one order of magnitude by 
employing laser beams which are ten times stronger than the current
value.

\begin{figure}
\begin{center}
\includegraphics[width=7cm]{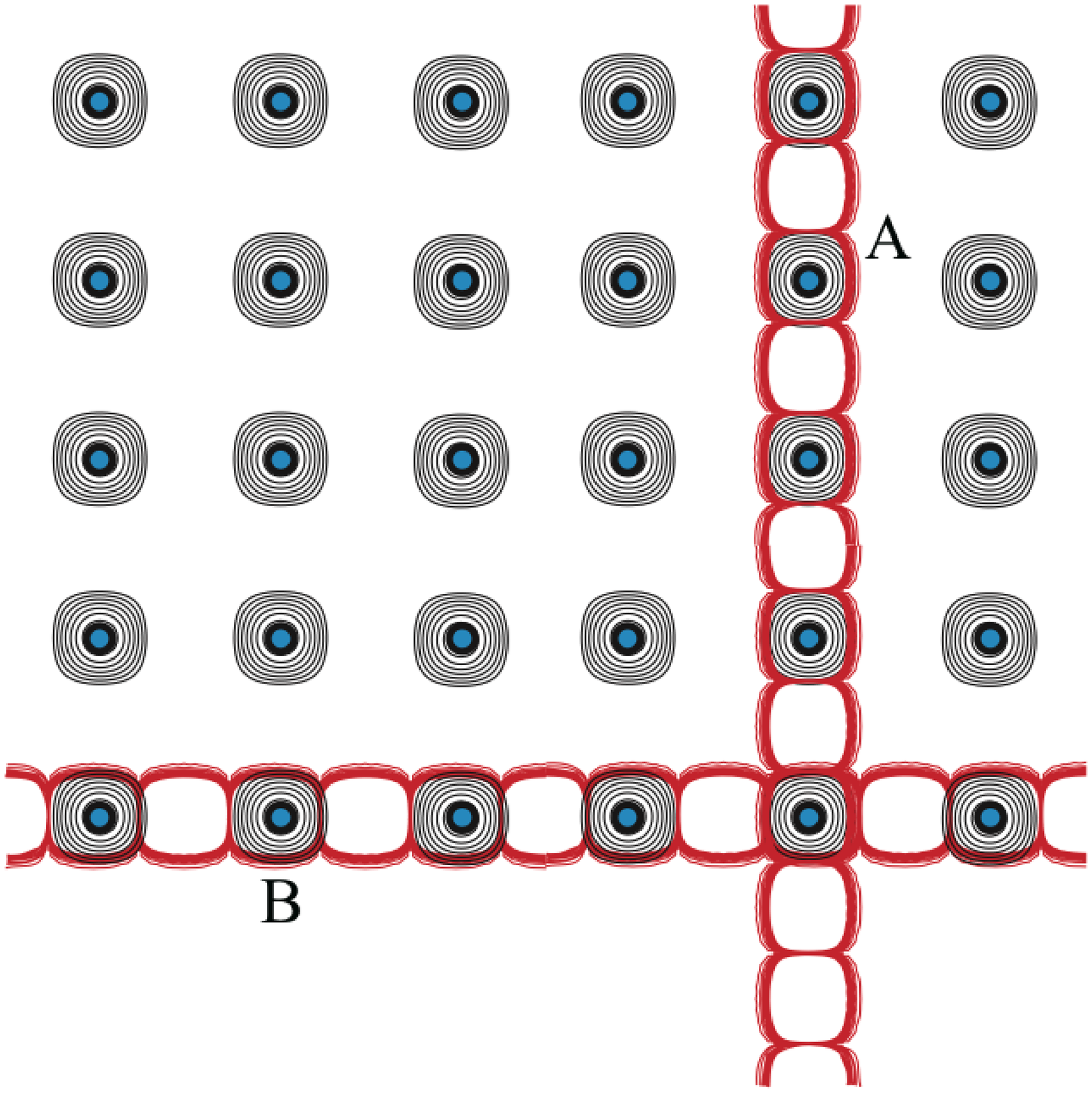}
\end{center}
\caption{(Color online) Two-qubit gate acts on atoms A and B 
in both cases. 
Two orthogonal optical lattices (light red contours) over two dimensional NFFD traps
(dark black contours). 
Atoms A and B are in arbitrary positions, not necessarily along one
of the primitive vectors of the array. The screen is looked down from the above.
\label{2d_optical_lattice}}
\end{figure}

In case the two atoms are in a general position, not necessarily
along a primitive lattice vector of the optical lattice, we may introduce 
two orthogonal optical lattices as shown in Fig. \ref{2d_optical_lattice} (a). 
Then the selected atomic
states $|0 \rangle$ of one atom and $|1 \rangle$ of the other 
meet at the intersection of the two optical lattices
to acquire the extra dynamical phase. It should be noted that 
two-qubit gate operations may be applied simultaneously and independently 
on many pairs of qubits.

\section{Conclusion}\label{summary}

We proposed a scalable design of a neutral atom quantum computer
with an on-demand interaction. A qubit is made of two hyperfine states of
an atom. Associated with each atom, there is an optical fiber, through
which a trap laser and a gate control laser are supplied.
The latter is used to implement one-qubit gates by two-photon
Raman transitions. 
We have shown that a selective two-qubit gate operation is possible
if atoms are trapped by an array of NFFD traps with a variable aperture size.
This would allow for true two-qubit gates, rather than performing the same 
gate operation with many atoms in massive parallelism as in the experiments 
by Mandel {\it et al} \cite{ref:mandel2}.
We have obtained an upper bound of the gate operation time of 7.87~ms
with the corresponding fidelity of 0.886. It may be possible to further
reduce the execution time or increase the fidelity by taking different
time-dependence of the control parameters. Reduction of the gate operation
time is also possible by increasing the laser intensity. 

We believe that most of our proposal can be demonstrated within current technology. 
The severe technical challenge is to attach the 
ends of the optical fibers to the apertures of the substrate. 
We strongly believe that the rapid progress 
in nanotechnology makes it possible to solve this in the foreseeable future.  
Some other issues, such as trapping atoms near material surfaces, 
how to form an optical lattice in close proximity to a surface, 
and how to load individual traps with single atoms, 
should be addressed seriously toward realization of our proposal. 
We plan to study these issues in the future and would like to encourage experimentalists 
to design experiments based on our proposal. 

\begin{acknowledgments}
One of the authors (M.N.) 
would like to thank Takuya Hirono for drawing his attention to
\cite{ref:sscat} and Yutaka Mizobuchi for discussions.
We are also grateful to Ken'ichi Nakagawa for enlightening discussions.
A part of this research is supported by
``Open Research Center''
Project for Private Universities: Matching fund subsidy from
MEXT (Ministry of Education, Culture, Sports, Science and
Technology).
K.K. is supported in part by Grant-in-Aid
for Scientific Research (Grant No.~21740267) from the MEXT of Japan. 
Y.K., M.N. and T.O. are partially supported by Grant-in-Aid for Scientific 
Research from the JSPS (Grant No.~23540470). 

\end{acknowledgments}





\appendix
\section{Numerical Simulations of Two-Qubit Gate Operation}\label{appsimulation}
The purpose of Appendix\ref{appsimulation} is to estimate the execution times 
of Steps 1 through 3 in the two-qubit gate operation introduced in Sec. \ref{twoqubitgate} 
by solving the time-dependent Schr\"{o}dinger equation 
\begin{equation}
i \hbar \frac{\partial \psi}{\partial t} =
-\frac{\hbar^2}{2m} \nabla^2 \psi + V \psi  \label{tdepseq}
\end{equation}
numerically. Using the experimentally relevant parameters, we calculate the optimized  
time to implement the gate operation keeping the fidelity 0.99 for each step. 
Here, the fidelity is defined as the overlap between the ground state 
wavefunction $\psi_{\rm fin}^{(n)} ({\bm x})$ at the final potential configuration 
in Steps 1, 2 and 3 and the solution of Eq. (\ref{tdepseq}): 
\begin{equation}
{\cal F} (t) = \int d {\bm x} \psi^{\ast}({\bm x},t) \psi_{\rm fin}^{(n)} ({\bm x}).
\end{equation}

\subsection{Step 1}\label{appstep1}

A red-detuned laser beam passing through an aperture with a radius on 
the order of the laser wavelength forms an attractive potential
$U_{\rm F}(\bm{x}, t)$ of Eq. (\ref{eq:rs}) to an atom. 
The potential profile is a function of time if the aperture radius changes 
as a function of time. 
We solve Eq. (\ref{tdepseq}) with $V=U_{\rm F}(\bm{x}, t)$ 
numerically to evaluate the time required to transfer an atom
from the vicinity of an aperture to a space point, which will
be a minimum of an optical lattice potential when it is turned on. 

We use the D$_1$ transition of $^{87}$Rb atom to trap an atom and
take the numerical values 
%
%
%
%
%
%
\begin{equation}
\begin{array}{ccc}
\mbox{length}&  \lambda_{\rm F}& 795.118~{\rm nm}\\
\mbox{energy}& U_0& 
\begin{array}{c}
k_B \times 20.1~\mu{\rm K} \\
= h \times 4.2 \times 10^5~{\rm Hz}
\end{array} \\
\mbox{time}& \tau_{\rm F}=\dfrac{\hbar}{U_0} &0.38~\mu{\rm s},
\end{array}
\label{parameterstep1}
\end{equation}
for the relevant parameters for definiteness, where the laser intensity is put to 
$I_0 = 10^5$~W/cm$^2$ \cite{Bandi}. Here $\lambda_{\rm F}$ is 
the wavelength of the trap laser and $U_0$ is the
potential depth. The wavelength $\lambda_{\rm F}$ is fixed through the detuning 
\begin{equation}
\Delta_{eg} \equiv \omega_{\rm L}-\omega_0=2\pi c \left( \frac{1}{\lambda_{\rm F}}
-\frac{1}{\lambda_0}\right) = -4.1\times 10^{11}~\mbox{Hz},
\end{equation}
where $\lambda_0 = 794.979$~nm is the wavelength of the D$_1$ 
transition \cite{ref:sscat}. With these parameters, the lifetime of atoms 
due to the photon recoil heating is given as $t_h \approx U_0 / 2 E_{F} \gamma_s \sim 1$ sec 
with the recoil energy $E_{F} = h^2/ 2 m \lambda_F^{2}$ and the photon 
scattering rate $\gamma_s = \Gamma_e \Omega_{eg}^2/ 4 \Delta_{eg}^2$, 
which is much longer than the total execution time.  

We rewrite the Schr\"odinger equation in a dimensionless form using the 
values given in (\ref{parameterstep1}), and introduce the cylindrical coordinates 
by taking advantage of the cylindrical symmetry around the $z$-axis. 
By writing $\phi = r \psi$,
the Schr\"odinger equation is put in the form
\begin{equation}\label{eq:sch1}
i \frac{\partial \tilde{\phi}}{\partial \tilde{t}} =
-\alpha \left[\frac{\partial^2 \tilde{\phi}}{\partial \tilde{r}^2}
-\frac{1}{\tilde{r}} \frac{\partial \tilde{\phi}}{\partial \tilde{r}}
+ \frac{\tilde{\phi}}{\tilde{r}^2}
+ \frac{\partial^2 \tilde{\phi}}{\partial \tilde{z}^2}
\right]  + \tilde{U}_{\rm F} \tilde{\phi},
\end{equation}
where dimensionless variables are denoted with a tilde
and $\alpha = \hbar^2/2m \lambda_{\rm F}^2 U_0 \simeq 2.2 \times 10^{-4}$.



We take the initial aperture radius
$a_{\rm ini} = 1.5~\lambda_{\rm F}$, corresponding to $\tilde{z}_{\rm ini} \simeq 2.0$,
 and the final aperture radius $a_{\rm fin}= 2.8~\lambda_{\rm F}$, for which
$\tilde{z}_{\rm fin} \simeq 8.0$, in our computation [see Fig. \ref{fig:nffd}]. 
Let $\tau^{(1)}$ be the dimensionless time required to change the aperture size
from $a_{\rm ini}$ to $a_{\rm fin}$. 
The aperture size is changed as
\begin{equation}
\tilde{a}(t) = a_{\rm ini} + (a_{\rm fin} - a_{\rm ini}) \sin^2\left( \frac{\pi}{2 \tilde{\tau}^{(1)}}\tilde{t} \right)
\ (0 \leq \tilde{t} \leq \tilde{\tau}^{(1)}).
\end{equation}
Here, we have vanishing initial and final velocities $d\tilde{a}/d\tilde{t}$
to avoid oscillatory behavior of the wave function during time evolution. 

\begin{figure}
\begin{center}
\includegraphics[width=8cm]{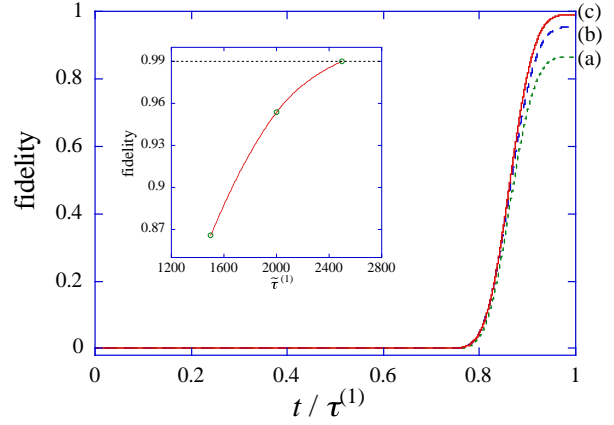}
\end{center}
\caption{(Color online) Time evolution of the fidelity ${\cal F}$ for $\tilde{\tau}^{(1)}=$ 1500 (a), 2000 (b), and 2500 (c).
Since the fidelity increases monotonically with $\tilde{\tau}^{(1)}$, 
the execution time is evaluated at which the fidelity reaches 0.99 by interpolating the data, 
as shown in the inset.}
\label{step1numerical}
\end{figure}
We solved the Schr\"odinger equation (\ref{eq:sch1}) numerically and found
$\tilde{\tau}^{(1)}= 2500$, that is, $\tau^{(1)} = \tilde{\tau}^{(1)} \tau_{\rm F}=
950~\mu$s, is required to attain the fidelity 0.99 [see Fig. \ref{step1numerical}].
This $\tau^{(1)}$ is consistent with the adiabaticity bound $\tau_{\rm ad}^{(1)}$ in (\ref{eq:tad1}).

\subsection{Step 2}\label{appstep2}
We switch from the NFFD potential 
$U_{\rm F}(\bm{x}, \tau^{(1)})$ to the optical lattice potential $U_{\rm OL}(\bm{x})$ 
with the time scale $\tau^{(2)}$ as
\begin{equation}
\begin{array}{c}
\displaystyle
V_{\rm F}(\bm{x}, t) = \cos^2 \left(\frac{\pi}{2 \tau^{(2)} } t\right) U_{\rm F}(\bm{x}, \tau^{(1)})
\vspace{.3cm}\\
\displaystyle
V_{\rm OL}(\bm{x}, t) = \sin^2 \left( \frac{\pi}{2 \tau^{(2)} } t \right) U_{\rm OL}(\bm{x})
\end{array}
\end{equation}
to this end. Here $U_{\rm F}(\bm{x}, \tau^{(1)})$ is the NFFD trap potential at $t=\tau^{(1)}$ and
\begin{equation}
U_{\rm OL}(\bm{x}) = -V_0 \cos^2(2 k_{\rm OL} x) e^{-2(y^2+(z-z_m)^2)/w^2}
\end{equation}
is the optical lattice potential, where $z_m$ is the $z$-coordinate of
the minimum of the NFFD potential at $t=\tau^{(1)}$. 

It turned out to be convenient to employ different scaling for physical
parameters here from those in Step~1. We use scales 
\begin{equation}
\begin{array}{ccc}
\mbox{length}&  \lambda_{\rm OL}& 785~{\rm nm}\\
\mbox{energy}& E_{\rm r} = \dfrac{\hbar^2k_{\rm OL}^2}{2m} & 
\begin{array}{c}
 k_{\rm B} \times 0.176~\mu{\rm K} \\ 
 = h \times 3.7 \times 10^3~{\rm Hz}
\end{array}\\
\mbox{time}& \tau_{\rm OL} =\dfrac{\hbar}{E_{\rm r}} &43~\mu{\rm s} 
\end{array}
\label{parameterstep2}
\end{equation}
and a dimensionless variable is denoted with a tilde as before.
%
%
%
%
%
%
%

Now the Schr\"odinger equation (\ref{tdepseq}) is put in a dimensionless form
as
\begin{eqnarray}
i \frac{\partial \tilde{\psi}}{\partial \tilde{t}} =
-\frac{1}{(2\pi)^2}  \tilde{\nabla}^2 \tilde{\psi}
+[\tilde{V}_{\rm OL}({\bm x},t)+\tilde{V}_{\rm F}({\bm x},t)] \tilde{\psi}
\label{eq:sch2}
\end{eqnarray}
For definiteness, we take $V_0/E_{\rm r} = 40, U_0/E_{\rm r}=114$ and $w/\lambda_{\rm OL}
= 4$. We have taken $V_0$ to be of the same order as
that in Mandel {\it et al} \cite{ref:mandel2}. It should be noted that there is no
rotational symmetry in the above Schr\"odinger equation and
it must be solved by using the Cartesian coordinate system.

We have solved the Schr\"odinger equation (\ref{eq:sch2}) numerically
and found that the time scale $\tau^{(2)} = 560~\mu$s is required to
attain the fidelity 0.99 with the similar analysis done in the Sec. \ref{appstep1}. 
Thus the above value of $\tau^{(2)}$ is consistent with this estimation.

\subsection{Step 3}
The optical transitions shows that
the effective potentials acting on $|0 \rangle=|F=1, m_F=1 \rangle$ 
and $|1 \rangle=|F=2, m_F=1 \rangle$ are
\begin{equation}
\begin{array}{c}
\displaystyle  V_{|0 \rangle}(x) = \frac{1}{4}V_+(x) + \frac{3}{4}V_-(x),
\vspace{.2cm}\\
\displaystyle  V_{|1 \rangle}(x) = \frac{3}{4}V_+(x) + \frac{1}{4}V_-(x),
\end{array}
\label{statedependpot}
\end{equation}
respectively. Here, we use the following decompositions
\begin{equation}
\begin{array}{c}
\displaystyle |0 \rangle = 
 -\frac{1}{2}\left|\frac{3}{2}, \frac{1}{2} \right\rangle  \left|\frac{1}{2}, 
\frac{1}{2} \right\rangle + 
\frac{\sqrt{3}}{2} \left|\frac{3}{2}, \frac{3}{2} \right \rangle
\left| \frac{1}{2}, -\frac{1}{2} \right\rangle,
\vspace{.2cm}\\ 
\displaystyle  |1 \rangle = 
\frac{\sqrt{3}}{2}\left|\frac{3}{2},\frac{1}{2} \right\rangle
\left|\frac{1}{2}, \frac{1}{2} \right\rangle+
\frac{1}{2}\left|\frac{3}{2}, \frac{3}{2}
\right\rangle
\left|\frac{1}{2}, -\frac{1}{2}\right\rangle, 
\end{array}
\end{equation}
where $|3/2, 1/2 \rangle|1/2,1/2 \rangle$ in the right hand side
denotes a vector with nuclear spin $(I = 3/2, I_z=1/2)$ 
and electron spin $(S=1/2, S_z=1/2)$, for example. 
If the polarization angle $\theta$ is increased, 
$V_{|0 \rangle}(x)$ ($V_{|1\rangle}(x)$) moves right (left).
The choice of these hyperfine states gives an exactly 
counter-moving potential Eq. (\ref{statedependpot}) 
for $| 0 \rangle$ and $| 1 \rangle$, 
while the states used by Mandel {\it et al}. \cite{ref:mandel2} 
yield the asymmetric potential given in Eq. (\ref{statepotMan}). 
Hence we restrict ourselves to the motion of $| 1 \rangle$ state below. 

\begin{figure}
\begin{center}
\includegraphics[width=8cm]{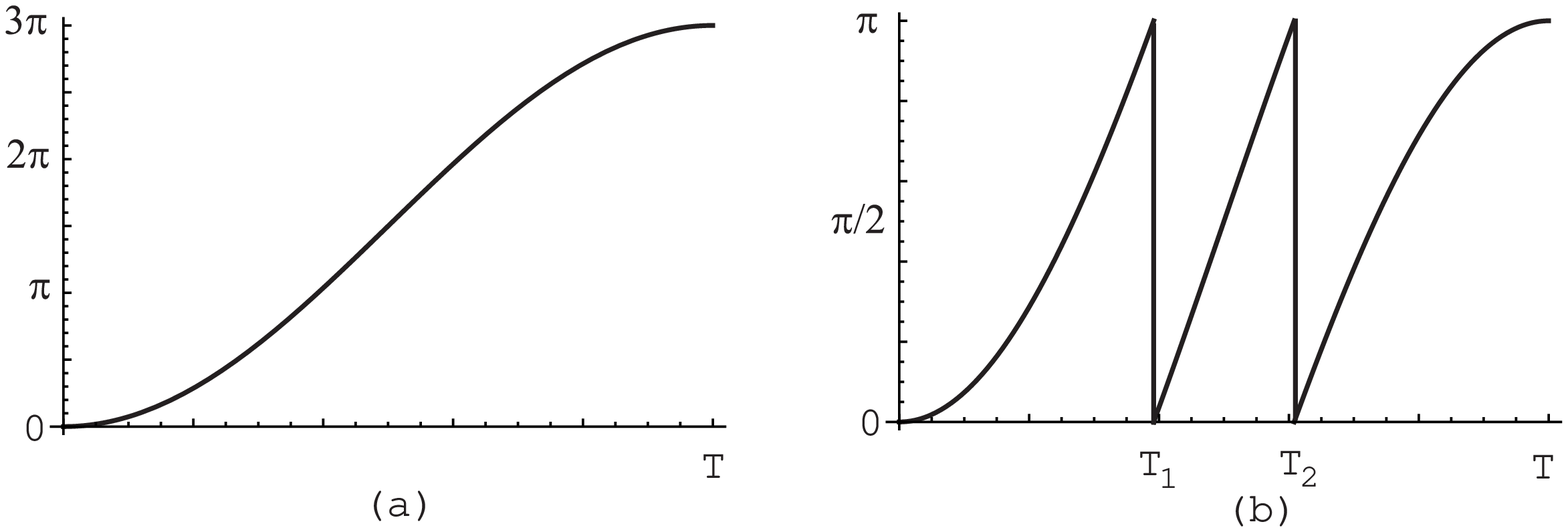}
\end{center}
\caption{(a) Polarization must be rotated by $3\pi$ to move an atom by
three wavelengths of the optical lattice. (b) Actual change of 
$\theta$ when EOM is in use. Even though $\theta$ has discontinuity at 
$t=T_1$ and $T_2$,
$\cos \theta$ is continuous in time. The effect of the angle $\theta$
shown in (b) on an atom is the same as that of $\theta$ in (a).
}
\label{fig:angle}
\end{figure}

We take the number of the lattice constant over which
an atom is transported to be $n$. 
In actual experiments, the angle $\theta$ may be changed by using
an electro-optic modulator (EOM). The angle $\theta$ is then
limited within the range $0\leq \theta \leq \pi$. This means
that we have to reset $\theta$ to zero as soon as it 
reaches to $\theta = \pi$, if we need to change $\theta$
beyond $\pi$ (Fig. \ref{fig:angle}).
The trick is that even when
$\theta$ jumps by $\pi$, $\cos \theta$ does not change at all.
This reset must be done instantaneously so that
the trapped atoms does not move in the optical lattice during the 
resetting time. By repeated use of this reset $n-1$ times, it is
possible to change $\theta$ effectively from $\theta=0$ to
$\theta = n \pi$.

We have solved the Schr\"odinger equation
\begin{equation}
i \hbar \frac{\partial \psi}{\partial t} =-\frac{\hbar^2}{2m}
\nabla^2 \psi + V_{|1 \rangle}(t) \psi
\end{equation}
numerically. The angle $\theta$ is changed as
\begin{equation}
\theta(t) = n \pi \sin^2 \left(\frac{\pi }{2\tau^{(3)} } t
\right),
\end{equation}
in case an atom is transferred in the optical lattice by $n$ site.
We employ the same scaling as introduced in Step 2, with which
the Sch\"ordinger equation is put into a dimensionless form as
\begin{equation}
i  \frac{\partial \tilde{\psi}}{\partial \tilde{t}} =
-\frac{1}{(2\pi)^2}\left(\frac{\partial^2 \tilde{\phi}}
{\partial \tilde{\rho}^2}
-\frac{1}{\tilde{\rho}} \frac{\partial \tilde{\phi}}{\partial \tilde{\rho}}
+ \frac{\tilde{\phi}}{\tilde{\rho}^2}
+ \frac{\partial^2 \tilde{\phi}}{\partial \tilde{z}^2}
\right)  + \tilde{V}_{|1 \rangle} \tilde{\phi}.
\end{equation}
Here, we use the cylindrical coordinates by taking advantage of the cylindrical 
symmetry of the optical lattice around the $z$-axis.

\begin{figure}
\begin{center}
\includegraphics[width=8cm]{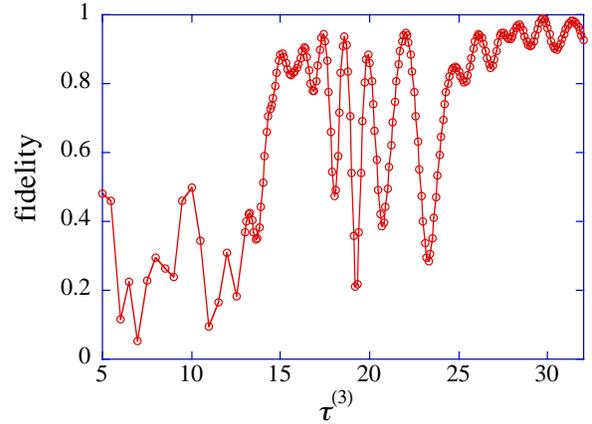}
\end{center}
\caption{(Color online) The fidelity as a function of $\tau^{(3)}$ for $n=6$.}
\label{fidelityn=62}
\end{figure}

The Schr\"odinger equation is solved numerically and we find that the
operation time $\tau^{(3)}$, which gives the fidelity 0.99, 
is 0.577~ms and 1.28~ms for two choices $n=3$ and $n=6$, respectively. 
Note that $\tau^{(3)}$ is not a monotonous function of $n$. This
is due to the fact that the lattice potentials $V_{|0 \rangle}$
and $V_{|1\rangle}$ are superpositions of two counterpropagating
plane waves $V_{\pm}$ and there is an interference between them.
In fact, the fidelity is an oscillating function of $\tau^{(3)}$
for a given $n$; an example is shown in Fig. \ref{fidelityn=62}. 
These complicated properties make the execution time $\tau^{(3)}$ 
several times longer than the time scale $\tau^{(3)}_{\rm ad}$ 
obtained from adiabaticity requirement in Sec. \ref{step3estimate}. 

%
%
%
%

\section{Equivalence of ${\cal F}_{0 \to T}$ and ${\cal F}_{T \to 0}$}\label{backward}
Let $| \psi_0 (0) \rangle $ and $| \psi_0 (T) \rangle $ be the eigenstates 
of the Hamiltonian at $t = 0$ and $t = T$, respectively. 
Under the potential change $V(t)$, and hence the
Hamiltonian change $H(t)$, the state develops as
\begin{equation}
| \psi (t) \rangle = U(t) | \psi_0 (0) \rangle, \hspace{3mm} U(t) = {\cal T} e^{-i \int_{0}^{t} H(t) dt}. 
\end{equation}
The fidelity of this process is defined as
\begin{equation}
{\cal F}_{0 \to T} = \langle \psi_{0}(T) | \psi(T) \rangle =  \langle \psi_{0}(T) |U(T)| \psi_{0} (0) \rangle. 
\end{equation}
For the reverse process $H(T) \to H(0)$, we start with $| \psi_0 (T) \rangle $. 
The time development operator is $\bar{U}(t) = U(t)^{-1}$. Then the fidelity is 
\begin{eqnarray}
{\cal F}_{T \to 0} &=& \langle \psi_{0}(0) |\bar{U}(T)| \psi_{0} (T) \rangle  \nonumber \\
&=& \langle \psi_{0}(0) |U^{-1}(T)| \psi_{0} (T) \rangle  \nonumber \\
&=& \langle \psi_{0}(T) |U(T)| \psi_{0} (0) \rangle^{\ast} \nonumber \\
&=& {\cal F}_{0 \to T}^{\ast} = {\cal F}_{0 \to T} ,
\end{eqnarray}
where the reality of ${\cal F}$ is used.




\end{document}